\documentclass[%
reprint,onecolumn,
 amsmath,amssymb
 aps,
]{revtex4-1}

\usepackage{graphicx}
\usepackage{dcolumn}
\usepackage{bm}
\usepackage[colorlinks, linkcolor=blue, anchorcolor=blue, citecolor=blue]{hyperref}
\usepackage{url}
\usepackage{amsmath,amssymb}
\usepackage{newtxmath}
\usepackage[mathlines]{lineno}
\usepackage{subfigure}
\usepackage{eurosym}
\usepackage{makecell}

\begin{document}

\preprint{APS/123-QED}

\title{Multiple electromagnetically induced transparency without a control field in an atomic array coupled to a waveguide}
\author{W. Z. Jia}
\email{wenzjia@swjtu.edu.cn}
\affiliation{School of Physical Science and Technology, Southwest Jiaotong University, Chengdu 610031, China}
\author{Q. Y. Cai}
\affiliation{School of Physical Science and Technology, Southwest Jiaotong University, Chengdu 610031, China}
\date{\today}

\begin{abstract}
We investigate multiple electromagnetically induced transparency (EIT) in a waveguide quantum electrodynamics (wQED) system containing 
an atom array. 
By analyzing the effective Hamiltonian of the system, we find that in terms of the single-excitation collective states, a properly designed $N$-atom 
array can be mapped into a driven ($N+1$)-level system that can produce multiple EIT-type phenomenon. The corresponding scattering spectra of 
the atom-array wQED system are discussed both in the single-photon sector and beyond the single-photon limit. 
The most significant feather of this type of EIT scheme is control-field-free, which may provide an alternative
way to produce EIT-like phenomenon in wQED system when external control fields are not available.
The results given in our paper may provide good guidance for future experiments on multiple EIT without a control field in wQED system.
\end{abstract}

\maketitle



\section{Introduction}
\label{intro}
Electromagnetically induced transparency (EIT) \cite{Harris-PRL1990,Boiler-PRL1991,Harris-PhysToday1997,Fleischhauer-RMP2005} is a phenomenon associated with destructive interference between 
two excitation pathways, where an otherwise opaque medium is rendered transparent to a resonant probe field due to the 
application of another strong control field. The phenomenon of EIT has many important applications including creation of  
slow light \cite{Hau-Nature1999,Liu-Nature2001} and production of giant nonlinear effects \cite{Schmidt-OL1996,Harris-PRL1998}, making it an important building block in quantum information and 
communication proposals. The early demonstrations of the EIT phenomenon 
are based on various three-level atomic vapors \cite{Harris-PRL1990,Boiler-PRL1991,Harris-PhysToday1997}. Further studies show that when a multi-level system pumped by more than one control fields, multiple transparency windows are created, allowing transmissions of multiple probe beams simultaneously at different wavelengths \cite{Paspalakis-PRA2002,Ye-PRA2002,Yelin-PRA2003,Goren-PRA2004,Zhang-PRL2007}. An typical example is an $(N+1)$-level atom with $N$ 
lower levels and a single upper level, where the $N-1$ metastable states are coupled near resonantly to the excited state 
by control fields, resulting in at most $N-1$ transparency windows occurring in the absorption spectrum for the probe field \cite{Paspalakis-PRA2002}.
Single- or multiple-window EIT or related phenomena have also been investigated in other systems, including rareearth-ion-doped crystals \cite{Ham-OptComm1997}, semiconductor 
quantum wells \cite{Serapiglia-PRL2000}, optical resonators \cite{Smith-PRA2004,Naweed-PRA2005,Xiao-PRA2007}, plasmonic resonator antennas
\cite{Kekatpure-PRL2010,Lu-PRA2012}, optomechanical systems \cite{Agarwal-PRA2010,Weis-Sci2010}, cavity magnomechanical systems \cite{Zhang-SciAdv2016,Ullah-PRA2020}, superconducting circuits \cite{Abdumalikov-PRL2010,Hoi-PRL2011,Novikov-nature2016,Long-PRL2018}, and so on. 

Recently, with the development of modern nanotechnology, waveguide quantum electrodynamics (wQED) structures \cite{Roy-RMP2017,Gu-PhysRep2017}, which are realized by strongly coupling a single atom or multiple atoms, to a one-dimensional (1D) waveguide, have brought about widespread attention. For their high atom-waveguide coupling efficiencies, the wQED systems become excellent platforms to manipulate transport of single or few photons \cite{Shen-OL2005,Shen-PRL2005,Chang-PRL2006,Shen-PRL2007,Shi-PRB2009,Shen-PRA2009,Astafiev-Sci2010,Longo-PRL2010,Zheng-PRA2010,Fan-PRA2010,Witthaut-NJP2010,Roy-PRL2011,Zheng-PRL2011,Liao-PRA2012,Jia-PRA2013,Laakso-PRL2014,Yang-Ann.Phys.2020,Cai-PRA2021}
, and may have potential applications in quantum devices at single-photon level \cite{Bermel-PRA2006,Chang-NPhys2007,Zhou-PRL2008,Aoki-PRL2009,Abdumalikov-PRL2010,Hoi-PRL2011,Bradford-PRL2012,Bradford-PRA2012,Hoi-PRL2013,Wang-PRA2014,Jia-PRA2017,Zhu-PRA2019}.
When multiple atoms are coupled to a 1D waveguide, the effective long-range interactions resulted from the photon exchanges, as well as the interferences between the reemitted photons from different atoms, can yield many interesting phenomena, such as superradiant and subradiant states \cite{Dicke-PR1954,Vetter-Phys.Scr.2016,Loo-science2013,Zhang-PRL2019,Ke-PRL2019,Wang-PRL2020,Dinc-PRR2019,Dinc-Quantum2019}, waveguide-mediated long-range entanglements between atoms \cite{Zheng-PRL2013,Ballestero-PRA2014,Facchi-PRA2016,Mirza-PRA2016}, creation of photonic band gap \cite{Fang-PRA2015,Greenberg-PRA2021}, micro cavity structures with atomic mirrors \cite{Chang-NJP2012,Mirhosseini-nature2019},  topology-enhanced nonreciprocal scattering \cite{Nie-PRApplied2021}, asymmetric Fano line shapes \cite{Tsoi-PRA2008,Cheng-OL2012,Liao-PRA2015,Cheng-PRA2017,Mukhopadhyay-PRA2019,Feng-PRA2021}, and so on. It is noteworthy that recent studies show that in wQED systems with double atoms, a new type of control-field-free EIT can be realized \cite{Shen-PRB2007,Fang-PRA2017,Ask-Arxiv2020}. In addition, single-window EIT-like phenomenon in a multi-atom wQED system was also studied \cite{Mukhopadhyay-PRA2020}.  For the scheme of control-field-free EIT, extra driving light fields are not required, which may provide alternative ways to produce EIT-type phenomenon in solid-state systems like superconducting circuits. The key to obtain EIT without a control field in wQED systems with two atoms is to generate dark (subradiant, decoupled from the waveguide) and bright (superradiant, coupled to the waveguide) modes and at the same time persist with the waveguide-induced interactions between them, making these collective states form an effective driven $\Lambda$-type atom \cite{Fang-PRA2017,Ask-Arxiv2020}.

Thus, a natural question is that whether this type of mapping can be generalized to the case of multi-atom wQED, and then be used to generate 
multi-window EIT. In this paper, by analyzing the effective Hamiltonian of system, we prove that if the separation between neighboring atoms is 
a half-integral multiple of the resonant wavelength, and the transition frequencies of atoms are different, there exist effective couplings between 
the superradiant state and the $N-1$ subradiant states [Fig.~\ref{system}(b)]. Thus in terms of these collective states, the system can be mapped to a driven $(N+1)$-level 
atom with $N$ lower levels and a single upper level [Fig.~\ref{system}(c)]. This configuration is exactly the one can exhibit multiple EIT, with at most $N-1$ transparency windows occurring in the system 
\cite{Paspalakis-PRA2002}.  As a verification, we further derive the analytic expressions of the scattering amplitudes of the atomic-chain wQED 
system under the EIT condition obtained from effective-Hamiltonian analysis. The results given in our paper may provide good guidance for 
future experiments on multiple EIT without a control field and have potential applications in multi-wavelength optical communication and 
quantum information processing. 

The paper is organized as follows. In Sec.~\ref {Model}, we give a theoretical model, obtain the EIT condition by analyzing the effective Hamiltonian, and further calculate the transmittance and reflectance of 
single-photons scattering. In Sec.~\ref {EIT-Spectra}, we analyze multiple EIT phenomenon in an atom array in detail. In Sec.~\ref {InelasticScatt}, 
we discuss the EIT-type scattering spectra beyond the single-photon limit. Finally, further discussions and conclusions are given 
in Sec.~\ref {conclusion}.

\section{\label{Model}Model}
\begin{figure*}
	\centering
	\includegraphics[width=0.8\textwidth]{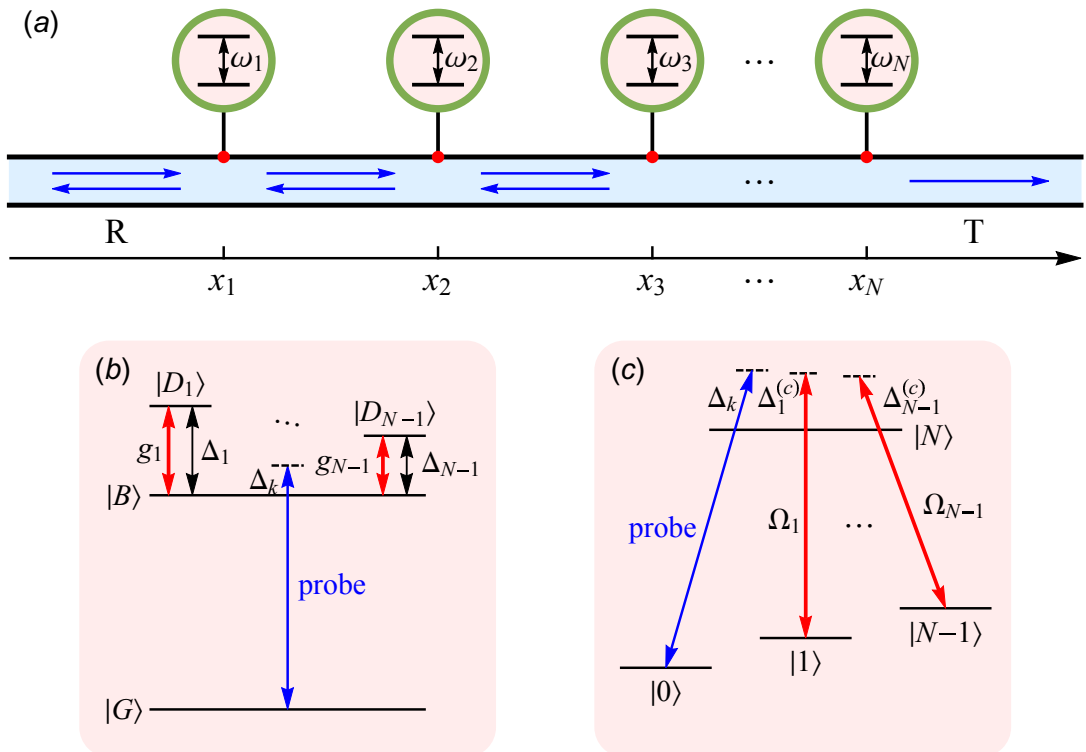}
	\caption{(a) Chain of $N$ two-level atoms coupled to a 1D waveguide. A probe field  is incident from the left. (b) Effective energy diagram of the atom array when restricted 
		to the single-excitation subspace, where $N-1$ subradiant states $|D_{i}\rangle$ ($i=1,2,\cdots N-1$) couple to the superradiant state  $|
		B\rangle$ with effective coupling strength $g_{i}$.  The frequency difference between 
		$|D_{i}\rangle$ and $|B\rangle$ is ${\rm\Delta}_{i}$. A weak probe is introduced via the waveguide to couple the transition between
		the ground state $|\mathrm{G}\rangle$ and  the superradiant state $|\mathrm{B}\rangle$. (c) Schematic diagram of a driven
		$(N+1)$-level atom. It consists of a ground state $|0\rangle$, $N-1$ metastable states $|i\rangle$ ($i=1,2,\cdots N-1$) and an excited 
		state $|N\rangle$. The metastable states are coupled near resonantly to the 
		excited state by $N-1$ laser fields with Rabi frequencies ${\rm\Omega}_i$ and detunings ${\rm\Delta}^{(\mathrm{c})}_{i}$. The transition between ground state and excited state is driven by a probe laser.}
	\label{system}       
\end{figure*}
We study $N$ periodically spaced two-level atoms coupled to photonic modes in a 1D waveguide with linear dispersion, as
shown schematically by Fig.~\ref{system}(a). The Hamiltonian of system can be written as ($\hbar=1$)
\begin{eqnarray}
\hat{H}&=&\sum_{i=1}^{N}\omega_i{\sigma}^{+}_{i}{\sigma}^{-}_{i}+\int\mathrm{d}x\hat{c}_\mathrm{R}^{\dagger}\left(x\right)\left(-\mathrm{i}v_{\mathrm{g}}\frac{\partial}{\partial x}\right)\hat{c}_\mathrm{R}\left(x\right)
+\int\mathrm{d}x\hat{c}_\mathrm{L}^{\dagger}\left(x\right)\left(\mathrm{i}v_{\mathrm{g}}\frac{\partial}{\partial x}\right)\hat{c}_\mathrm{L}\left(x\right)
\nonumber \\ 
&&+\sum_{i=1}^{N}\int\mathrm{d}xV_{i}\delta\left(x-x_i\right)\left\{\left[\hat{c}_\mathrm{R}^{\dagger}\left(x\right)+\hat{c}_\mathrm{L}^{\dagger}\left(x\right)\right]{\sigma}^{-}_{i}+\mathrm{H}.\mathrm{c}.\right\}.
\label{Hamitonian}
\end{eqnarray}
Here $\omega_i$ represents the transition frequency of the $i$th atom. ${\sigma}^{+}_{i}=|\mathrm{e}\rangle_{i}\langle \mathrm{g}|$ and 
${\sigma}^{-}_{i}=|\mathrm{g}\rangle_{i}\langle \mathrm{e}|$ are the raising and lowering operators of the $i$th atom, 
where $|\mathrm{e}\rangle_{i} (|\mathrm{g}\rangle_{i})$ represents the corresponding excited (ground) state. $v_\mathrm{g}$ is the group velocity of the photons in the waveguide. 
$\hat{c}_\mathrm{R}(x)$ [$\hat{c}^\dagger_\mathrm{R}(x)$] and $\hat{c}_\mathrm{L}(x)$ [$\hat{c}^\dagger_\mathrm{L}(x)$] 
are the field operators of annihilating (creating) the right- and left-propagating photons at position $x$ in the waveguide. $V_i$ is the coupling strength of the $i$th atom at the position $x_i$. 

\subsection{\label{EffectiveH}Effective-Hamiltonian analysis}
It is instructive to establish the mapping between the collective states of an atom array and the levels of a single driven $(N+1)$-level atom, which can help us to understand the physical mechanism of the EIT-like phenomenon in wQED system with many atoms. We assume that the frequency differences between different atoms are small. After tracing out the photon modes in the waveguide and neglecting the non-Markovian effects, we can obtain the effective non-Hermitian Hamiltonian of atom array \cite{Zhang-PRL2019,Ke-PRL2019} 
\begin{equation}
\hat{H}_{\mathrm{eff}}=\sum_{i=1}^{N}\left(\overline{\omega}+\delta\omega_i\right)\sigma_{i}^{+}\sigma_{i}^{-}-\frac{\mathrm{i}}{2}\sum_{i,j=1}^{N}\sqrt{{\rm\Gamma}_i{\rm\Gamma}_j}e^{\mathrm{i}\left|\phi_i-\phi_j\right|}\sigma_{i}^{+}\sigma_{j}^{-},
\label{Heff1}
\end{equation}
where $\overline{\omega}=(\sum_{i=1}^{N}\omega_i)/N$ is the average frequency of the atoms for reference, $\delta\omega_i=\omega_i-\overline{\omega}$ is the detuning between the atomic transition frequency and the reference frequency, ${\rm\Gamma}_i=2V_i^2/v_\mathrm{g}$  is the decay rate of a single atom coupled to the waveguide, $\phi_{i}=\overline{\omega}x_i/v_\mathrm{g}$ is the phase acquired by a photon with  frequency $\overline\omega$ traveling from the origin to the coupling point $x_i$. The off diagonal elements of this effective Hamiltonian describe both coherent and dissipative atom-atom interactions mediated by the waveguide modes. 

Here we focus on the case that the decay rates are the same ${\rm\Gamma}_i={\rm\Gamma}$, the transition frequencies are not necessarily equal but $\delta\omega_i\ll\overline{\omega}$ is satisfied, and the phase delay between neighboring atoms is a constant $\phi_{i+1}-\phi_{i}=n\pi$ ($n\in\mathbb{N}^{+}$), i.e., the separation between neighboring atoms is a half-integral multiple of the resonant wavelength. The corresponding effective Hamiltonian becomes
\begin{equation}
\hat{H}_{\mathrm{eff}}=\hat{H}^{(0)}_{\mathrm{eff}}+\sum_{i=1}^{N}\delta\omega_i{\sigma}^{+}_{i}{\sigma}^{-}_{i},
\label{Heff2}
\end{equation}
with
\begin{eqnarray}
\hat{H}^{(0)}_{\mathrm{eff}}=\sum_{i=1}^{N}\overline{\omega}{\sigma}^{+}_{i}{\sigma}^{-}_{i}
-\mathrm{i}\frac{{\rm\Gamma}}{2}\sum_{i,j=1}^{N}\left (-1\right)^{\left(i-j\right)n}\sigma_{i}^{+}\sigma_{j}^{-}.
\label{Heff2_0}
\end{eqnarray}
Note that one can obtain a diagonalized Hamiltonian
\begin{equation}
\hat{H}^{(0)}_{\mathrm{eff}}=\sum_{i=1}^{N}\left(\overline{\omega}-\frac{\mathrm{i}}{2}N{\rm\Gamma}\delta_{1i}\right)\tilde{\sigma}_{i}^{+}\tilde{\sigma}_{i}^{-}
\label{Heff20D}
\end{equation}
by introducing the collective atomic raising and lowering operators 
\begin{equation}
\tilde{\sigma}_{i}^{\pm}=\frac{1}{\sqrt{N}}\sum_{j=1}^{N}e^{\pm \mathrm{i}\frac{2\pi}{N}\left(i-1\right)j}\left(-1\right)^{\left(j-1\right)n}\sigma_{j}^{\pm}.
\label{collective}
\end{equation}
Clearly, there are $N$ eigenstates of $\hat{H}^{(0)}_{\mathrm{eff}}$ in the single-excitation sector, among which one state $|\mathrm{B}\rangle={\tilde{\sigma}}_{1}^{+}|\mathrm{G}\rangle$ is a superradiant state with enhanced decay rate $N{\rm\Gamma}$, and the other $N-1$ states $|\tilde{\mathrm{D}}_i\rangle={\tilde{\sigma}}_{i+1}^{+}|\mathrm{G}\rangle ~(i=1,2\cdots N-1$) are subradiant ones with zero decay rate \cite{Vetter-Phys.Scr.2016,Wang-PRL2020}. Here $|\mathrm{G}\rangle=|\mathrm{g}\rangle_1|\mathrm{g}\rangle_2\cdots|\mathrm{g}\rangle_N$ is the ground state of the atom array. One can rewrite the effective Hamiltonian \eqref{Heff2} in terms of collective atomic operators as
\begin{eqnarray}
\hat{H}_{\mathrm{eff}}&=&
\hat{H}_{\mathrm{B}}+\hat{H}_{\mathrm{D}}+\hat{H}_{\mathrm{BD}},
\label{Heff3}
\end{eqnarray}
where
\begin{subequations}
	\begin{eqnarray}
	\hat{H}_{\mathrm{B}}=\left(\overline{\omega}-\frac{\mathrm{i}}{2}N{\rm\Gamma}\right)\tilde{\sigma}_{1}^{+}\tilde{\sigma}_{1}^{-}
	\label{HB},
	\end{eqnarray}
	\begin{eqnarray}
	\hat{H}_{\mathrm{D}}=\sum_{i=2}^{N}\overline{\omega}\tilde{\sigma}_{i}^{+}\tilde{\sigma}_{i}^{-}+\sum_{\substack{i,j=2\\i\neq j}}^{N}\left(\tilde{g}_{ij}\tilde{\sigma}_{i}^{+}\tilde{\sigma}_{j}^{-}+\mathrm{H}.\mathrm{c}\right),
	\label{HD}
	\end{eqnarray}
	\begin{eqnarray}
	\hat{H}_{\mathrm{BD}}=\sum_{i=2}^{N}\left(\tilde{g}_{1i}\tilde{\sigma}_{1}^{+}\tilde{\sigma}_{i}^{-}+\mathrm{H}.\mathrm{c}\right)
	\label{HBD}.
	\end{eqnarray}
\end{subequations}
The coupling strengths between the collective modes can be written as 
\begin{equation}
\tilde{g}_{ij}=\frac{1}{N}\sum_{m=1}^{N}\delta\omega_m e^{\mathrm{i}\frac{2\pi}{N}\left(j-i\right)m}.
\label{gij}
\end{equation}
One can see that for $N$ identical atoms with $\delta\omega_i=0$, the coupling strengths $\tilde{g}_{ij}$ become vanish, which means that 
the collective states are decoupled from each other. On the contrary, when the transition frequencies of the atoms are different, 
i.e., $\delta\omega_i\neq 0$, there exists coherent interactions between these states, which plays a fundamental role to create EIT-type 
phenomena in an atom array. 

One can always find a unitary transformation $\sigma_{\mathrm{D}_{i}}^{-}=\sum_{j=1}^{N-1}{\mathbf{V}}_{ij}\tilde{\sigma}_{j+1}^{-}$ ($i=1,2,\cdots,N-1$) to diagonalize the effective Hamiltonian \eqref{HD} in the subradiant subspace. The unitarity condition requests  $\sum_{j=1}^{N-1}{\mathbf{V}}^{*}_{ij}{\mathbf{V}}_{mj}=\delta_{im}$. 
Clearly, the eigenstate $|\mathrm{D}_i\rangle=\sigma_{\mathrm{D}_{i}}^{+}|\mathrm{G}\rangle=\sum_{j=1}^{N-1}{\mathbf{V}}^{*}_{ij}|\tilde{\mathrm{D}}_j\rangle$ of Hamiltonian \eqref{HD} is also subradiant.
And we relabel $\tilde{\sigma}_{1}^{\pm}$ as  $\sigma_{\mathrm{B}}^{\pm}$. Then the effective Hamiltonian can be rewritten as 
\begin{eqnarray}
\hat{H}_{\mathrm{eff}}=
\left(\overline{\omega}-\frac{\mathrm{i}}{2}N{\rm\Gamma}\right)\sigma_{\mathrm{B}}^{+}\sigma_{\mathrm{B}}^{-}
+\sum_{i=1}^{N-1}\left(\overline{\omega}+{\rm\Delta}_{i}\right)\sigma_{\mathrm{D}_{i}}^{+}\sigma_{\mathrm{D}_{i}}^{-}
+\sum_{i=1}^{N-1}\left(g_i\sigma_{\mathrm{B}}^{+}\sigma_{\mathrm{D}_{i}}^{-}+\mathrm{H}.\mathrm{c}.\right),
\label{Heff4}
\end{eqnarray}
with effective detuning
\begin{eqnarray}
{\rm\Delta}_{i}=\sum_{\substack{j,m=1\\j\neq m}}^{N-1}\tilde{g}_{j+1,m+1}{\mathbf{V}}_{ij}{\mathbf{V}}^{*}_{im}
\label{effdetuning}
\end{eqnarray}
and effective coupling strength
\begin{equation}
g_{i}=\sum_{j=1}^{N-1}\tilde{g}_{1,j+1}{\mathbf{V}}^{*}_{ij}.
\label{effcoupling}
\end{equation}
When dealing with the single-photon scattering problem, we should only consider the ground state $|\mathrm{G}\rangle$ and the single-excitation 
states $|\mathrm{B}\rangle$ and  $|\mathrm{D}_i\rangle$. One can see from Eq.~\eqref{Heff4} that, the transition between the ground state 
$|\mathrm{G}\rangle$ and the surperradiant state $|\mathrm{B}\rangle$ is coupled by the waveguide modes with decay rate $N{\rm\Gamma}$. Thus a weak probe can be introduced via the waveguide to couple the transition $|\mathrm{G}\rangle\leftrightarrow|\mathrm{B}\rangle$ with a drving term ${\rm\Omega}_{\mathrm{p}}\sigma_{\mathrm{B}}^{+}+\mathrm{H}.\mathrm{c}.$. 
Moreover, the surperradiant state $|\mathrm{B}\rangle$ and the subradiant state $|\mathrm{D}_i\rangle$ are coherently coupled to each 
other with strength $g_{i}$, while the 
subradiant states are decoupled from each other. The corresponding energy diagram is shown in Fig.~\ref{system} (b). Clearly, in terms 
of these collective states, the system can be mapped to an $(N+1)$-level atom that can exhibit multiple EIT \cite{Paspalakis-PRA2002}, where 
the transition between the ground state $|0\rangle$ and the excited state $|N\rangle$ is coupled by the waveguide modes, and the 
metastable state $|i\rangle$ ($i=1,2,\cdots N-1$) is coupled near-resonantly to the excited state $|N\rangle$ by control fields with Rabi 
frequencies ${\rm\Omega}_{i}$ ($i=1,2,\cdots N-1$), as shown in Fig.~\ref{system} (c). To see this mapping more clearly, we derive the effective 
Hamiltonian of an $(N+1)$-level atom coupled to a waveguide in Appendix \ref{Hamiltonian-multi-levelatom}. By comparing the Hamiltonian 
\eqref{Heff4} with the Hamiltonian \eqref{HeffAtom}, we can make the identifications $|\mathrm{G}\rangle\leftrightarrow|0\rangle$, 
$|\mathrm{B}\rangle\leftrightarrow|N\rangle$, $|\mathrm{D}_{i}\rangle\leftrightarrow|i\rangle$, $\overline{\omega}\leftrightarrow\tilde\omega_{N}$, $g_{i}\leftrightarrow{\rm\Omega}_{i}$, 
${\rm\Delta}_{i}\leftrightarrow{\rm\Delta}^{(\mathrm{c})}_{i}$ and $N{\rm\Gamma}\leftrightarrow{\rm\Gamma}_{N0}$. 

Thus, like a driven $(N+1)$-level system, the atom array can exhibit multiple EIT for a single photon traveling in the waveguide, with at 
most $N-1$ transparency windows appearing at detuning ${\rm\Delta}_i$. This requires that all the atomic frequencies are different. If the 
atomic frequencies are equally spaced (at intervals of ${\rm\Delta}$) and the number of atom is small with $N=2-6$, the expressions of the effective 
detunings ${\rm\Delta}_i$ and effective coupling strengths $g_i$ can be calculated analytically by using Eqs.(\ref{effdetuning}) and (\ref{effcoupling}), which are summarized in Table~\ref{table}. Note that 
for the special case of two atoms $N=2$, the corresponding results agree with those in Refs.~\cite{Fang-PRA2017,Ask-Arxiv2020}.
Note that the multiple EIT scheme discussed here is control-field-free. Namely, the effective couplings between the collective excitations are mediated by the waveguide modes, thus external driving fields are not required, which is very different from the usual EIT phenomenon in a multi-level quantum system (e.g., a $\Lambda$-type atom). 

If the frequencies of some atoms are equal, the number of transparency windows will decrease. Specifically, if there are $m_i$ ($i=1,2\cdots M$) different atoms all with the same frequencies, and the other $m_0$ atoms are nonidentical, satisfying $\sum_{i=0}^{M}m_i=N$. One can prove that in this case, each type of $m_i$ identical atoms as a whole can be looked on as a single atom with effective decay $m_i {\rm\Gamma}$,  thus the system forms an effective array containing $m_0+M$ emitters, and the number of transparency windows decrease to $m_0+M-1$ (see Appendix~\ref{DegenerateCase}). 
\begin{table*}[t]
	\centering\caption{The expressions of the effective detunings ${\rm\Delta}_i$ and effective coupling strengths $g_i$ when the atomic frequencies are equally spaced at intervals of ${\rm\Delta}$. Without loss of generality, we set ${\rm\Delta}>0$.}
	\label{table}
	\begin{tabular}{cccccc}
		\hline\noalign{\smallskip}
		&$N$&$\delta\omega_j$&Effective detuning&Effective coupling\\
		\noalign{\smallskip}\toprule\noalign{\smallskip}
		&\makecell*{2}&\makecell*{$\pm\frac{{\rm\Delta}}{2}$}
		&\makecell*{${\rm\Delta}_{1}=0$}&\makecell*{$|g_{1}|=\frac{{\rm\Delta}}{2}$}&\\
		
		&\makecell*{3}&\makecell*{$\pm{\rm\Delta},~0$}
		&\makecell*{${\rm\Delta}_{1}=-{\rm\Delta}_{2}=\frac{\sqrt{3}}{3}{\rm\Delta}$}
		&\makecell*{$|g_{1}|=|g_{2}|=\frac{\sqrt{3}}{3}{\rm\Delta}$}\\
		
		&\makecell*{4}&\makecell*{$\pm\frac{3{\rm\Delta}}{2},~\pm\frac{{\rm\Delta}}{2}$}
		&\makecell*{${\rm\Delta}_{1}=-{\rm\Delta}_{3}=\frac{\sqrt{5}}{2}{\rm\Delta}$,~${\rm\Delta}_2=0$}
		&\makecell*{$|g_{1}|=|g_{3}|=\frac{\sqrt{10}}{5}{\rm\Delta}$,~$|g_{2}|=\frac{3\sqrt{5}}{10}{\rm\Delta}$}\\
		
		&\makecell*{5}&$\pm2{\rm\Delta},~\pm{\rm\Delta},~0,$&\makecell*{${\rm\Delta}_{1}=-{\rm\Delta}_{4}=\sqrt{\frac{15-\sqrt{145}}{10}}{\rm\Delta}$,\\ ${\rm\Delta}_2=-{\rm\Delta}_{3}=\sqrt{\frac{15+\sqrt{145}}{10}}{\rm\Delta}$} &\makecell*{$|g_{1}|=|g_{4}|=\sqrt{\frac{145+\sqrt{145}}{290}}{\rm\Delta}$,\\$|g_{2}|=|g_{3}|=\sqrt{\frac{145-\sqrt{145}}{290}}{\rm\Delta}$} \\
		
		&\makecell*{6}&$\pm\frac{5{\rm\Delta}}{2},~\pm\frac{3{\rm\Delta}}{2},~\pm\frac{{\rm\Delta}}{2}$
		&\makecell*{${\rm\Delta}_{1}=-{\rm\Delta}_{5}=\sqrt{\frac{35-8\sqrt{7}}{12}}{\rm\Delta}$,\\ ${\rm\Delta}_2=-{\rm\Delta}_{4}=\sqrt{\frac{35+8\sqrt{7}}{12}}{\rm\Delta}$,\\${\rm\Delta}_{3}=0$}
		&\makecell*{$|g_{1}|=|g_{5}|=\sqrt{\frac{440+16\sqrt{7}}{777}}{\rm\Delta}$,\\$|g_{2}|=|g_{4}|=\sqrt{\frac{440-16\sqrt{7}}{777}}{\rm\Delta}$,\\$|g_{3}|=\sqrt{\frac{675}{1036}}{\rm\Delta}$}  \\
		\noalign{\smallskip}\hline	
	\end{tabular}
\end{table*}
\subsection{\label{SAmplitudes}Expressions of the scattering amplitudes}
In previous subsection, we have mapped the collective states of the two-level atom array into a driven $(N+1)$-level system. To verify this 
analysis, starting from the full atom-waveguide Hamiltonian \eqref{Hamitonian}, we will solve the single-photon scattering problem, and 
provide the analytic expressions of the scattering amplitudes of the atom-array wQED system. Further analysis on the EIT-type spectra will be 
provided in the next section. We assume that initially a single photon with energy $E$ incidences. Thus, in the single excitation subspace, the 
eigenstate of the system can be written as 
\begin{eqnarray}
\left|\Psi\right>=
\int\mathrm{d}x\Phi_\mathrm{R}\left(x\right)\hat{c}^\dagger_\mathrm{R}\left(x\right)\left|\emptyset\right>
+\int\mathrm{d}x\Phi_\mathrm{L}\left(x\right)\hat{c}^\dagger_\mathrm{L}\left(x\right)\left|\emptyset\right>+\sum_{i=1}^{N}f_i{\sigma}^{+}_{i}\left|\emptyset\right>,
\label{eigenstate}
\end{eqnarray}
where $\Phi_\mathrm{R}(x) [\Phi_\mathrm{L}(x)]$ is the single-photon wave function of a right-moving (left-moving) photon. 
$f_i$ is the excitation amplitude of the $i$th atom. $|\emptyset\rangle$ is the vacuum state, 
which means that there are no photons in the waveguide and all atoms are in their ground states. Substituting Eq.~\eqref{eigenstate} into the eigen equation
\begin{equation}
\hat{H}\left|\Psi\right>=E\left|\Psi\right>
\label{eigenequation}
\end{equation}
yields the following equations of motion:
\begin{subequations}
	\begin{equation}
	\left(-\mathrm{i}v_\mathrm{g}\frac{\partial}{\partial x}-E\right)\Phi_\mathrm{R}\left(x\right)+\sum_{i=1}^{N}V_{i}\delta\left(x-x_{i}\right)f_i=0,
	\label{EQMA1}
	\end{equation}
	\begin{equation}
	\left(\mathrm{i}v_\mathrm{g}\frac{\partial}{\partial x}-E\right)\Phi_\mathrm{L}\left(x\right)+\sum_{i=1}^{N}V_{i}\delta\left(x-x_{i}\right)f_i=0,
	\label{EQMA2}
	\end{equation}
	\begin{equation}
	\sum_{i=1}^{N}V_{i}\left[\Phi_\mathrm{R}\left(x_{i}\right)+\Phi_\mathrm{L}\left(x_{i}\right)\right]+\left(\omega_i-E\right)f_i=0.
	\label{EQMA3}
	\end{equation}
\end{subequations}
Assuming that the photon is incident from the left, $\Phi_\mathrm{R}(x)$ and $\Phi_\mathrm{L}(x)$ take the form 
\begin{subequations}
	\begin{eqnarray}
	\Phi_\mathrm{R}\left(x\right)=
	e^{\mathrm{i}kx}\Big[\theta\left(x_{1}-x\right)+\sum_{i=1}^{N-1}t_{i}\theta\left(x-x_{i}\right)\theta\left(x_{i+1}-x\right)
	+t\theta\left(x-x_{N}\right)\Big],
	\label{PhiR}
	\end{eqnarray}
	\begin{equation}
	\Phi_\mathrm{L}\left(x\right)=
	e^{-\mathrm{i}kx}\Big[r\theta\left(x_{1}-x\right)+\sum_{i=1}^{N-1}r_{i}\theta\left(x-x_{i}\right)\theta\left(x_{i+1}-x\right)\Big],
	\label{PhiL}
	\end{equation}
\end{subequations}
where $k$ is the wave vector of the photon, $t_i$ ($r_i$) is the transmission (reflection) amplitude for the $i$th [$(i+1)$th] coupling point, 
$t$ ($r$) is the transmission (reflection) amplitude for the last (first) coupling point, 
and $\theta(x)$ denotes the Heaviside step function. Substituting Eqs.~\eqref{PhiR} and~\eqref{PhiL} into Eqs.~\eqref{EQMA1} -~\eqref{EQMA3}, we can fix $E=v_{\mathrm{g}}k$ and obtain
\begin{subequations}
	\begin{eqnarray}
	t_{i}=t_{i-1}-\mathrm{i}\frac{V_i}{v_{\mathrm{g}}}f_{i}e^{-\mathrm{i}\phi_{i}},
	\label{tj}
	\end{eqnarray}
	\begin{eqnarray}
	r_{i-1}=r_{i}-\mathrm{i}\frac{V_i}{v_{\mathrm{g}}}f_{i}e^{\mathrm{i}\phi_{i}},
	\label{rj}
	\end{eqnarray}
	\begin{eqnarray}
	f_{i}=\frac{\mathrm{i}V_i}{2\left({\rm\Delta}_{k}-\delta\omega_{i}\right)}\left[\left(t_{i}
	+t_{i-1}\right)e^{\mathrm{i}\phi_{i}}+\left(r_{i}+r_{i-1}\right)e^{-\mathrm{i}\phi_{i}}\right],
	\label{fj}
	\end{eqnarray}
\end{subequations}
with ${\rm\Delta}_{k}=v_{\mathrm{g}}k-\overline{\omega}$ being the detuning between the frequency of incident photon and the average frequency of atoms. The phase factor $\phi_{i}=\overline{\omega}x_i/v_\mathrm{g}$ is defined the same as that in previous subsection. 
Note that in this definition, we have made the Markov approximation by replacing the wave vector $k$ by $\overline{\omega}/v_\mathrm{g}$. 
Substituting $f_i$ from Eq.~\eqref{fj} into \eqref{tj} and \eqref{rj}, we obtain a recursive linear matrix equation,
\begin{eqnarray}
\begin{pmatrix}
t_{i}\\
r_{i}
\end{pmatrix}=
\mathbf{T}^{-1}_{\phi_i}\mathbf{T}_i\mathbf{T}_{\phi_i}
\begin{pmatrix}
t_{i-1}\\
r_{i-1}
\end{pmatrix},
\label{matrixeq}
\end{eqnarray}
with
\begin{subequations}
	\begin{eqnarray}
	\mathbf{T}_i=
	\begin{pmatrix}
	2-\alpha_i & 1-\alpha_i\\
	-1+\alpha_i & \alpha_i
	\end{pmatrix},
	\label{Tj}
	\end{eqnarray}
	\begin{eqnarray}
	\mathbf{T}_{\phi_i}=\begin{pmatrix}
	e^{\mathrm{i}\phi_{i}} & 0\\
	0 & e^{-\mathrm{i}\phi_{i}}
	\end{pmatrix}.
	\label{Tphi}
	\end{eqnarray}
\end{subequations}
Here $\alpha_i=\left({\rm\Delta}_{k}-\delta\omega_{i}+\mathrm{i}{\rm\Gamma}_i/2\right)/\left({\rm\Delta}_{k}-\delta\omega_{i}\right)$. According to
the analysis in Sec.~\ref{EffectiveH}, we assume that the separation between neighboring atoms is a half-integral multiple of the resonant wavelength, i.e., $\phi_{i+1}-\phi_{i}=n\pi$ ($n\in\mathbb{N}^{+}$),
and the atom-waveguide decay rates are equal, with ${\rm\Gamma}_i={\rm\Gamma}$, so that
the atom array can form an effective $(N+1)$-level systems like Fig.~\ref{system} (b). The boundary conditions require $t_0 = 1$, $t_N = t$, $r_N = 0$, and $r_0 =r$. Starting from the relation \eqref{matrixeq}, after iterative calculation, we obtain the following connective relation between the reflection and transmission amplitudes
\begin{eqnarray}
\begin{pmatrix}
t\\
0
\end{pmatrix}=
\prod_{i=1}^{N}\mathbf{T}_i
\begin{pmatrix}
1\\
r
\end{pmatrix}
\label{matrixres}
\end{eqnarray}
After some simplifications, we can obtain the 
expressions of transmission and reflection amplitudes 
\begin{subequations}
	\begin{equation}
	t=\frac{1}{1+\mathrm{i}\frac{{\rm\Gamma}}{2}\sum_{i=1}^{N}{\left({\rm\Delta}_{k}-\delta\omega_{i}\right)}^{-1}},
	\label{transmission1}
	\end{equation}
	\begin{equation}
	r=\frac{\mathrm{i}\frac{{\rm\Gamma}}{2}\sum_{i=1}^{N}{\left({\rm\Delta}_{k}-\delta\omega_{i}\right)^{-1}}}{1+\mathrm{i}\frac{{\rm\Gamma}}{2}\sum_{i=1}^{N}{\left({\rm\Delta}_{k}-\delta\omega_{i}\right)}^{-1}}.
	\label{reflection1}
	\end{equation}
\end{subequations}
One can further define the transmittance $T=|t|^2$ and the reflectance $R=|r|^2$. Note that the transmittance and the reflectance are constrained by the relation $T+R=1$ because of conservation of photon number. Thus in the following part, we focus on the reflectance $R$ only.
\section{\label{EIT-Spectra}SINGLE-PHOTON  EIT-LIKE SPECTRA}
\begin{figure*}
	\centering
	\includegraphics[width=1\textwidth]{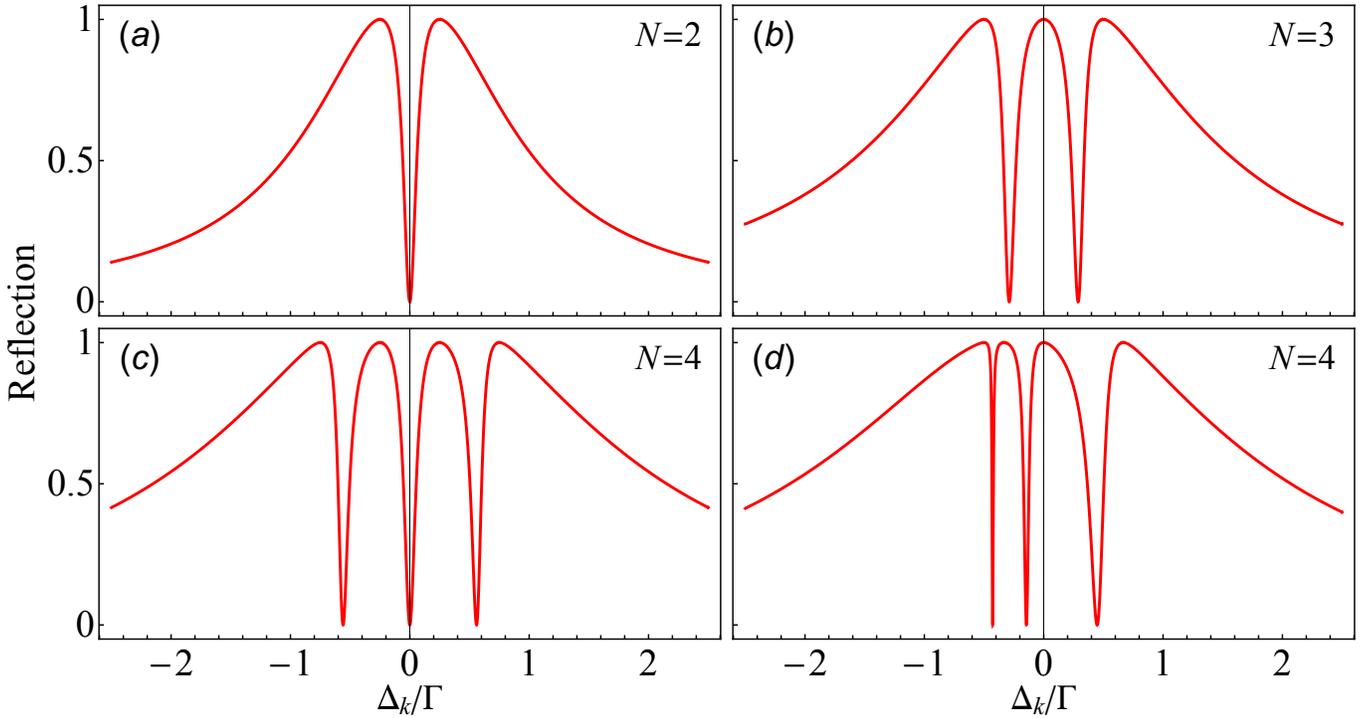}
	\caption{The reflection spectra for an array of $N$ atoms, with parameters
		(a) $N=2$, $\delta\omega_{1}=-{\rm\Gamma}/4$, $\delta\omega_{2}={\rm\Gamma}/4$; 
		(b) $N=3$, $\delta\omega_{1}=-{\rm\Gamma}/2$, $\delta\omega_{2}=0$, $\delta\omega_{3}={\rm\Gamma}/2$; 
		(c) $N=4$, $\delta\omega_{1}=-3{\rm\Gamma}/4$, $\delta\omega_{2}=-{\rm\Gamma}/4$, $\delta\omega_{3}={\rm\Gamma}/4$, $\delta\omega_{4}=3{\rm\Gamma}/4$; 
		(d) $N=4$, $\delta\omega_{1}=-{\rm\Gamma}/2$, $\delta\omega_{2}=-{\rm\Gamma}/3$, $\delta\omega_{3}=0$, $\delta\omega_{4}=2{\rm\Gamma}/3$.
	}
	\label{nondegenerate}
\end{figure*}

\begin{figure*}
	\centering
	\includegraphics[width=1\textwidth]{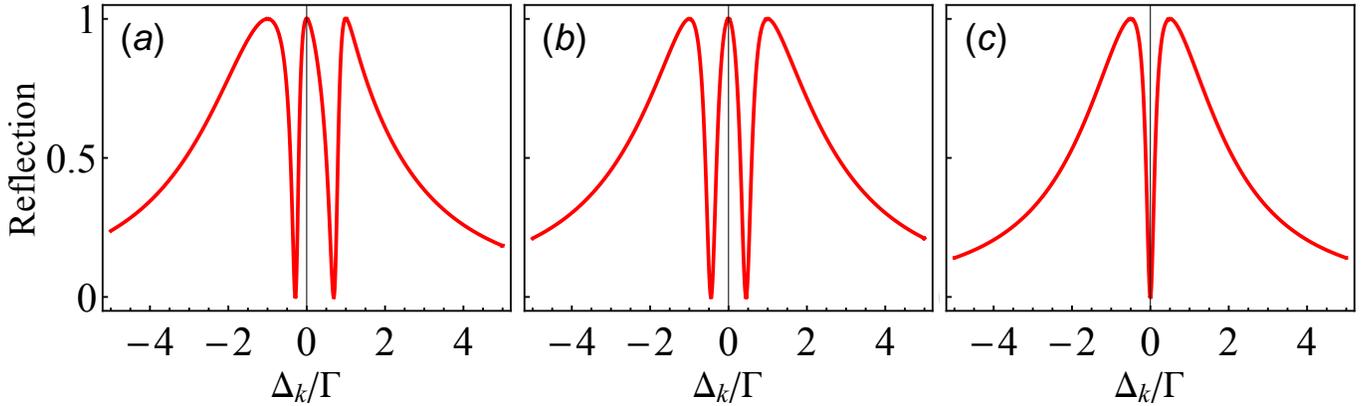}
	\caption{The reflection spectra for an array of $5$ atoms, with parameters
		(a) $\delta\omega_{1}=\delta\omega_{2}=\delta\omega_{3}=-{\rm\Gamma}$, $\delta\omega_{4}=0$, $\delta\omega_{5}={\rm\Gamma}$; 
		(b) $\delta\omega_{1}=\delta\omega_{2}=-{\rm\Gamma}$, $\delta\omega_{3}=0$, $\delta\omega_{4}=\delta\omega_{5}={\rm\Gamma}$; 
		(c) $\delta\omega_{1}=\delta\omega_{2}=-{\rm\Gamma}/2$, $\delta\omega_{3}=\delta\omega_{4}={\rm\Gamma}/2$;}
	\label{degenerate}
\end{figure*}

Based on the expression \eqref{reflection1}, we give a few examples of EIT-type reflection spectra that could occur in an atom array, as shown in Fig.~\ref{nondegenerate} and Fig.\ref{degenerate}. One can check that these EIT-type spectra are characterized by the parameters (${\rm\Delta}_i$ and $g_i$) of the effective control fields defined in Eq.~\eqref{effdetuning} and Eq.~\eqref{effcoupling}. 

When all the atomic frequencies are different, we can obtain EIT-type spectra containing $N-1$ total transparency points appearing at ${\rm\Delta}_k={\rm\Delta}_i$ ($i=1,2,\cdots N-1$). Specifically, in Figs.~\ref{nondegenerate}(a)-\ref{nondegenerate}(c), the atom number is $N=2$, $N=3$, and $N=4$, respectively. And the atomic frequencies
are equally spaced between $-(N-1){\rm\Delta}/2$ and $(N-1){\rm\Delta}/2$ at intervals of ${\rm\Delta}={\rm\Gamma}/2$. 
The corresponding spectra exhibit symmetric multiple EIT phenomenon. For these cases, the parameters of the effective control fields are summarized in Table~\ref{table}. 
In Fig.~\ref{nondegenerate}(d), we provide an example of the atomic frequencies
being unequally spaced, where the spectrum exhibits asymmetric EIT. 

We also give some examples with several atomic frequencies being equal [Figs.~\ref{degenerate}(a)-\ref{degenerate}(c)]. As analyzed in Sec.~\ref{EffectiveH} and Appendix~\ref{DegenerateCase}, each type of $m_i$ ($i=1,2,\cdots M$) identical atoms as a subsystem can be looked on as a single atom with effective decay $m_i {\rm\Gamma}$.  These $M$ collective states together with the other $m_0$ atoms, make the $N$-atom array be reduced to an effective array containing $m_0+M$ emitters, resulting in $m_0+M-1$ transparency windows, as shown in Fig.~\ref{degenerate}(a)-\ref{degenerate}(c). Note that the case shown in Fig.~\ref{degenerate}(c), where a single-window EIT-like spectrum is produced by utilizing two type of identical atoms, was also studied in \cite{Mukhopadhyay-PRA2020}. 

Note that in Figs.~\ref{nondegenerate} and \ref{degenerate}, the parameters have been appropriately chosen to ensure that 
the transparency windows are EIT type. If some of the frequency differences between the atoms are larger than the atomic line width, the 
relevant spectra will exhibit character of Autler-Townes splitting (ATS). Thus it is an important issue to determine whether the transparency 
windows in a spectrum are consequences of EIT or ATS \cite{Abi-Salloum-PRA2010,Anisimov-PRL2011}.
To this end, we decompose the reflection amplitude Eq.~\ref{reflection1} into the sum of several terms $r=\sum_{i=1}^{N}\tilde{r}_i$, where
\begin{equation}
\tilde{r}_i=\frac{A_{i}}{{\rm\Delta}_{k}-Z_{i}}
\label{ri}
\end{equation}
are Lorentz-type amplitudes, where $Z_{i}$ are the complex roots of the denominator of the scattering amplitude. 
The real and imaginary parts of $Z_{i}$ correspond to the resonance point and the half-width of the $i$th resonance, respectively.  
For large atom number, although it is hard to find analytic expressions of $A_{i}$ and $Z_{i}$, but they can be calculated numerically. 
By analyzing these roots, we can determine the type of transparency windows. 

\begin{figure*}
	\centering\includegraphics[width=1\textwidth]{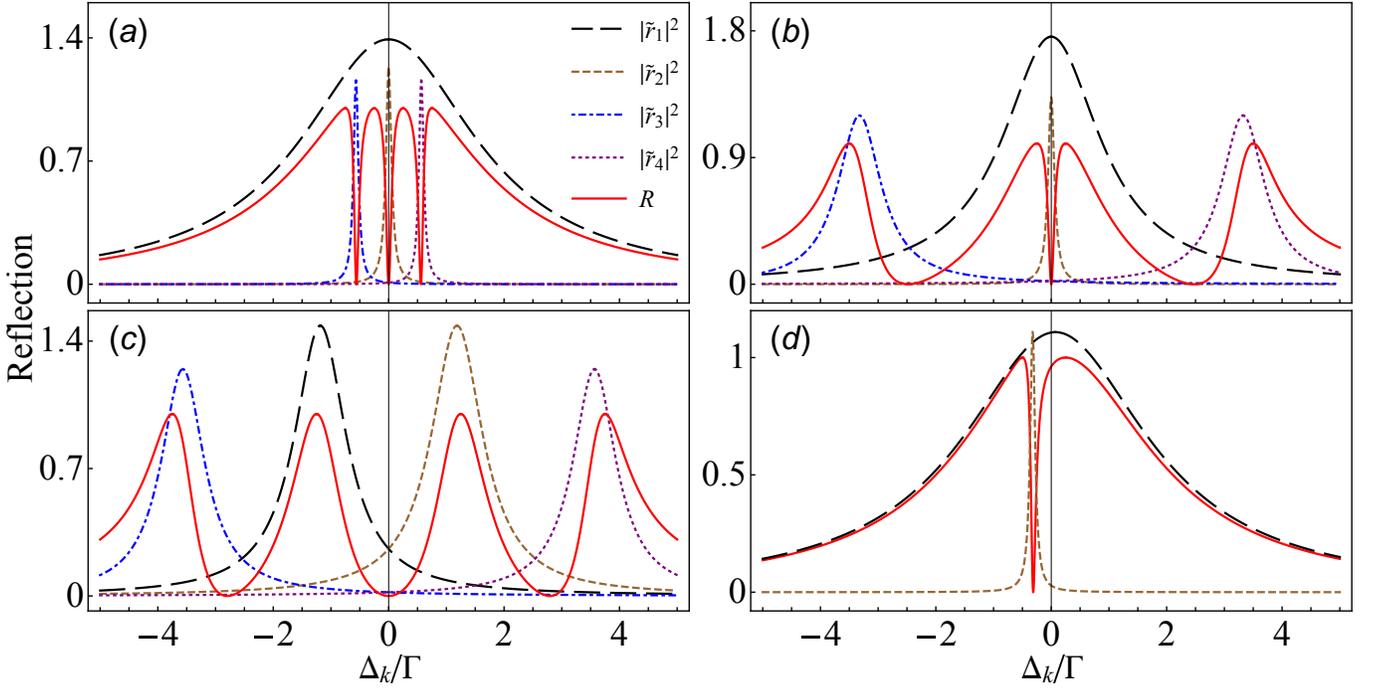}
	\caption{Reflection spectrum ($R$) and the different contributions to it ($\tilde{r}_i$) under different frequency differences between atoms. The atom number
		is $N=4$ for all cases. 
		(a) All the transparency windows are EIT-type, with parameters $-\delta\omega_{1}=\delta\omega_{4}=3{\rm\Gamma}/4$, $-\delta\omega_{2}=\delta\omega_{3}={\rm\Gamma}/4$;		
		(b) The EIT- and ATS-type windows coexist, with parameters $-\delta\omega_{1}=\delta\omega_{4}=7{\rm\Gamma}/2$, $-\delta\omega_{2}=\delta\omega_{3}={\rm\Gamma}/4$;	
		(c) All the transparency windows are ATS-type, with parameters $-\delta\omega_{1}=\delta\omega_{4}=15{\rm\Gamma}/4$, $-\delta\omega_{2}=\delta\omega_{3}=5{\rm\Gamma}/4$;	
		(d) EIT-type spectrum when some of the atomic frequencies are identical, with parameters
		$\delta\omega_{1}=-{\rm\Gamma}/2$, $\delta\omega_{2}=\delta\omega_{3}=\omega_{4}={\rm\Gamma}/4$.}
	\label{EITandAT}
\end{figure*}

To show this, we take the case of $N=4$ as an example,  
and provide the reflection coefficient as well as the resonances contained in it under different parameter regimes, as shown in 
Fig.~\ref{EITandAT}. Specifically, Fig.~\ref{EITandAT}(a) shows the case that all the three transparency windows are EIT-type. The 
atomic frequencies are chosen as $-\delta\omega_{1}=\delta\omega_{4}=3{\rm\Gamma}/4$ and $-\delta\omega_{2}=\delta\omega_{3}={\rm\Gamma}/4$. Under these parameters, the complex roots of the denominator of the scattering amplitude are $Z_{1}/{\rm\Gamma}\approx-1.839\mathrm{i}$,  $Z_{2}/{\rm\Gamma}\approx-0.059\mathrm{i}$, and $Z_{3,4}/{\rm\Gamma}\approx\mp0.566-0.051\mathrm{i}$, respectively, corresponding to three narrow resonances at ${\rm\Delta}_k=0,\pm0.566{\rm\Gamma}$ and a wide resonance at ${\rm\Delta}_k=0$ [see the long dashed, the short dashed, the dot-dashed and the dotted lines in Fig.~\ref{EITandAT}(a)].
The Fano-type destructive interference between the wide and the narrow resonances produces a reflection spectrum with three EIT-type transparency points located at ${\rm\Delta}_k=0,\pm0.566{\rm\Gamma}$, as shown by the solid line in Fig.~\ref{EITandAT}(a).
In Fig.~\ref{EITandAT}(b), we show the case that the EIT- and ATS-type windows coexist. The atomic frequencies are chosen as  $-\delta\omega_{1}=\delta\omega_{4}=7{\rm\Gamma}/2$ and $-\delta\omega_{2}=\delta\omega_{3}={\rm\Gamma}/4$. Correspondingly, the complex roots are $Z_{1}/{\rm\Gamma}\approx-1.022\mathrm{i}$, $Z_{2}/{\rm\Gamma}\approx-0.067\mathrm{i}$, and $Z_{3,4}/{\rm\Gamma}\approx\mp 3.32-0.456\mathrm{i}$, respectively. According to these results, we can see that the destructive interference between the narrow and the wide resonances at ${\rm\Delta}_k=0$ can create an EIT-type transparency point. On the other hand, the distance between the left (right) and the central resonances is larger than their width, thus the observed dip can be interpreted as a gap between the two peaks, thus the left and the right windows are ATS-type, as shown in Fig.~\ref{EITandAT}(b).
Fig.~\ref{EITandAT}(c) shows the case that all the three transparency windows are ATS-type. The atomic frequencies are chosen as  $-\delta\omega_{1}=\delta\omega_{4}=15{\rm\Gamma}/4$ and $-\delta\omega_{2}=\delta\omega_{3}=5{\rm\Gamma}/4$. In this regime, we have $Z_{1,2}/{\rm\Gamma}\approx\mp1.184-0.544\mathrm{i}$, $Z_{3,4}/{\rm\Gamma}\approx\mp3.567-0.456\mathrm{i}$.
We can see that in this case the distances between any pair of neighboring resonances are larger than their width, thus all the windows in the reflection spectrum are ATS-type, where the dips can be interpreted as gaps between neighboring resonances,  as shown in Figs.~\ref{EITandAT}(c).
Finally, Fig.~\ref{EITandAT}(d) shows a case that some of the atomic frequencies are identical, with $\delta\omega_{1}=-{\rm\Gamma}/2$, $\delta\omega_{2}=\delta\omega_{3}=\delta\omega_{4}={\rm\Gamma}/4$. In this case, there are two poles of the reflection amplitude, $Z_{1}/{\rm\Gamma}=0.073-1.948\mathrm{i}$, $Z_{2}/{\rm\Gamma}=-0.323-0.052\mathrm{i}$, corresponding to a narrow resonances at ${\rm\Delta}_k=-0.323{\rm\Gamma}$ and a wide resonance at ${\rm\Delta}_k=0.073{\rm\Gamma}$. Consequently, there exists only one EIT-type dip located at ${\rm\Delta}_{k}=-0.323{\rm\Gamma}$, caused by destructive interference between two resonances, as shown in Fig.~\ref{EITandAT}(d). We can see that in this case the number of transparency windows decreases, as discussed in Sec.~\ref{EffectiveH} and Appendix.~\ref{DegenerateCase}. 
\section{\label{InelasticScatt}SCATTERING SPECTRA BEYOND THE SINGLE-PHOTON LIMIT}
\begin{figure*}
	\centering\includegraphics[width=0.8\textwidth]{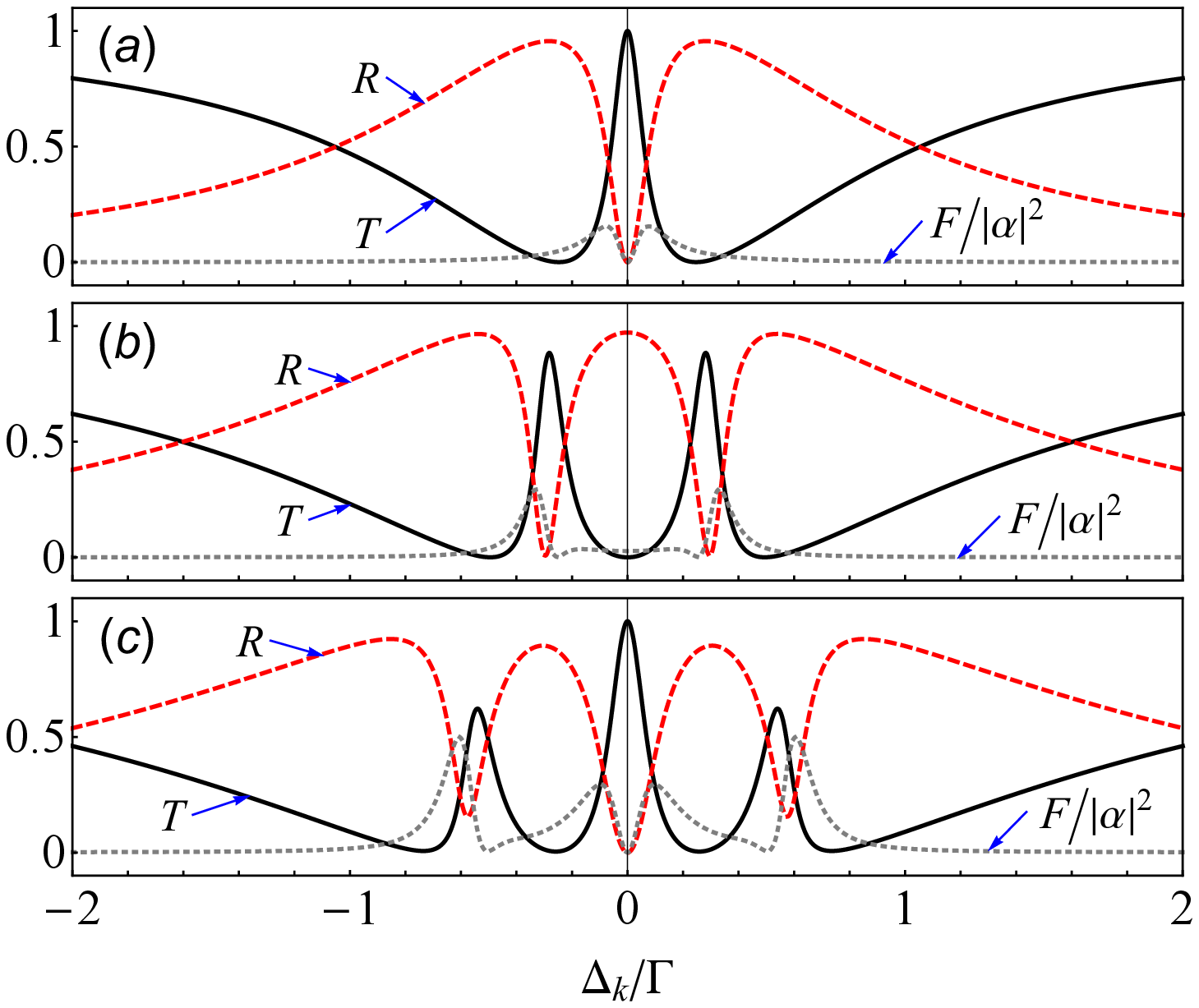}
	\caption{Transmission coefficient (solid lines), reflection coefficient (dashed lines), and inelastic photon flux (dashed lines) as a function of probe detuning ${\rm\Delta}_{k}$ for different atom number. 
		(a) $N=2$, $-\delta\omega_{1}=\delta\omega_{2}={\rm\Gamma}/4$, the intensity of the coherent drive is $|\alpha|^2=0.01{\rm\Gamma}$; 
		(b) $N=3$, $-\delta\omega_{1}=\delta\omega_{3}={\rm\Gamma}/2$, $\delta\omega_{2}=0$, the coherent drive amplitude is $|\alpha|^2=0.01{\rm\Gamma}$; 
		(c) $N=4$, $-\delta\omega_{1}=\delta\omega_{4}=3{\rm\Gamma}/4$, $-\delta\omega_{2}=\delta\omega_{3}={\rm\Gamma}/4$, the coherent drive amplitude is $|\alpha|^2=0.04{\rm\Gamma}$.}
	\label{inelastic}
\end{figure*}
In previous sections, we have showed that in the one-photon sector, the atom array can exhibit EIT-type spectra because the ground state  and the single-excitation states of system can form an effective $(N+1)$-level systems like Fig.~\ref{system} (b). 
However, if the probe field is a coherent field containing some multi-photon components, the multiply excited states may be occupied, making the setup not an effective $(N+1)$-level systems. In this case, if the steady state of the driven system contains the ingredient of single-excited superradiant state, the transmittance will decrease because of inelastic scattering.  On the contrary, if the system can evolve into a dark steady state being orthogonal to single-excited superradiant state, the total transparency point can preserves even beyond the one-photon sector.  To verify these results, we calculate numerically the scattering coefficients and the total inelastic photon flux. In our simulation we use a weak coherent field, which can include some multi-photon components, as a probe. In a frame rotating with the drive frequency $\nu=v_{\mathrm{g}}k$, the master equation for the driven atom array can be written as \cite{Ask-Arxiv2020,Kockum-PRL2018}
\begin{eqnarray}
\dot{\hat{\rho}}=
-\mathrm{i}\left[{\hat{H}_{\mathrm{drive}},\hat{\rho}}\right]+\sum_{i=1}^{N}{\rm\Gamma}_{i}\mathcal{D}\left[{\sigma}_{i}^{-}\right]\hat{\rho}
+\sum_{i\neq j}{\rm\Gamma}_{ij}\left({\sigma}_{i}^{-}\hat{\rho}{\sigma}_{j}^{+}-\frac{1}{2}\{{\sigma}_{i}^{+}{\sigma}_{j}^{-},\hat{\rho}\}\right),
\label{master eq}
\end{eqnarray}
with
\begin{eqnarray}
\hat{H}_{\mathrm{drive}}=-\sum_{i=1}^{N}\left({\rm\Delta}_{k}-\delta\omega_{i}\right){\sigma}_{i}^{+}{\sigma}_{i}^{-}+\sum_{i\neq j}\mathcal{G}_{ij}{\sigma}_{i}^{+}{\sigma}_{j}^{-}
+\sum_{i=1}^{N}\left({\rm\Omega}_{i}e^{\mathrm{i}(\phi_{i}-\phi_{1})}{\sigma}_{i}^{+}+\mathrm{H.c.}\right),
\label{Hdrive}
\end{eqnarray}
where $\mathcal{G}_{ij}=\frac{1}{2}\sqrt{{\rm\Gamma}_i{\rm\Gamma}_j}\sin|\phi_i-\phi_j|$ is the exchange interaction between the atoms mediated by the waveguide modes, ${\rm\Gamma}_{ij}=\sqrt{{\rm\Gamma}_i{\rm\Gamma}_j}\cos|\phi_i-\phi_j|$ is the collective decay. $\mathcal{D}[\hat{O}]\hat{\rho}=\hat{O}\hat{\rho}\hat{O}^{\dag}-\{\hat{O}^{\dag}\hat{O},\hat{\rho}\}/2$ is the Lindblad operator.  ${\rm\Omega}_{i}=\sqrt{\frac{{\rm\Gamma}_{i}}{2}}\alpha$ is the Rabi frequency of the atom $i$, and $|\alpha|^{2}$ is the number of photons per second coming from the coherent drive. The
other qualities  are the same as those defined in defined in Sec.~\ref{Model}.   

Using input-output theory, the transmission and reflection amplitudes can be defined as \cite{Ask-Arxiv2020}
\begin{subequations}
	\begin{equation}
	t=\frac{\langle\hat{b}^{(\mathrm{t})}_{\mathrm{out}}\rangle}{\alpha}=e^{i(\phi_{N}-\phi_{1})}-\mathrm{i}\frac{1}{\alpha}\sum_{i=1}^{N}e^{i(\phi_{N}-\phi_{i})}\sqrt{\frac{{\rm\Gamma}_{i}}{2}}\langle{\sigma}^{-}_{i}\rangle,
	\end{equation}
	\begin{equation}
	r=\frac{\langle\hat{b}^{(\mathrm{r})}_{\mathrm{out}}\rangle}{\alpha}=-\mathrm{i}\frac{1}{\alpha}\sum_{i=1}^{N}e^{\mathrm{i}(\phi_{i}-\phi_1)}\sqrt{\frac{{\rm\Gamma}_{i}}{2}}\langle{\sigma}^{-}_{i}\rangle,
	\end{equation}
\end{subequations}
where $\hat{b}^{(\mathrm{t})}_{\mathrm{out}}$ and $\hat{b}^{(\mathrm{r})}_{\mathrm{out}}$ are output operators describing the transmission and reflection light fields, $\langle{\sigma}^{-}_{i}\rangle=\mathrm{Tr}[\hat{\rho}{\sigma}^{-}_{i}]$ is the steady-state expectation value of lower operator ${\sigma}^{-}_{i}$, which can be obtained by numerically solving the master equation \eqref{master eq}. The corresponding transmission
and reflection coefficients are $T=|t|^2$ and $R=|r|^2$.

When the system is driven by a coherent field with frequency $\nu$, the inelastic power spectra of the output fields are defined as 
\begin{equation}
S_{\nu}^{(i)}(\omega)=\int{e^{-\mathrm{i}\omega t}}\langle\hat{b}_{\mathrm{out}}^{(i)\dag}(t)\hat{b}_{\mathrm{out}}^{(i)}(0)\rangle\mathrm{d}t.
\end{equation} 
Here $i=\mathrm{t},\mathrm{r}$ is used to label the transmitted and reflected fields, respectively. $\langle\hat{b}_{\mathrm{out}}^{(i)\dag}(t)\hat{b}_{\mathrm{out}}^{(i)}(0)\rangle$ is the steady-state correlation function, which can be calculated using the solution to the master equation \eqref{master eq}. 
The total inelastic photon flux can be further defined as \cite{Fang-PRA2015} 
\begin{equation}
F(\nu)=
\sum_{i=\mathrm{t},\mathrm{r}}\int S_{\nu}^{(i)}(\omega)\mathrm{d}\omega.
\end{equation} 

We plot the transmittance $T$, the reflectance $R$, and the inelastic photon flux $F$ as functions of probe detuning ${\rm\Delta}_{k}$ in the EIT regime in 
Figs~\ref{inelastic}(a)-\ref{inelastic}(c). The atom number is $2$, $3$, and $4$, respectively. The results show that these quantities satisfy relation $F/|\alpha|^2=1-T-R$, showing that photon-number conservation is preserved. And the atomic frequencies
are equally spaced between $-(N-1){\rm\Delta}/2$ and $(N-1){\rm\Delta}/2$ at intervals of ${\rm\Delta}={\rm\Gamma}/2$. The phase delay between neighboring atoms is set as $\pi$. The coupling strengths between the atoms and the waveguide are equal, with ${\rm\Gamma}_i={\rm\Gamma}$ and ${\rm\Omega}_i={\rm\Omega}$.  Without loss of generality, we assume the Rabi frequency ${\rm\Omega}$ is real. When a coherent driving field containing multiple-photon components incident, the multiply excited states may be occupied, making the setup not an effective $(N+1)$-level systems. Typically, when the atom number $N$ is an even number, the transparency point at ${\rm\Delta}_k=0$ is still hold, corresponding to a dark steady state of the system, which is an eigenstate of $\hat{H}_{\mathrm{drive}}$ with eigenvalue zero. Specifically, when $N=2$, the analytic expressions of the dark steady states are 
\begin{equation}
|\mathcal{D}_2\rangle=\frac{1}{\mathcal{N}_2}\left[{\rm\Delta}|\mathrm{gg}\rangle+2{\rm\Omega}\left(|\mathrm{eg}\rangle+|\mathrm{ge}\rangle\right)\right]
\end{equation}
with $\mathcal{N}_2=\sqrt{8{\rm\Omega}^2+{\rm\Delta}^2}$ being the normalization constant. Clearly, the dark steady state $|\mathcal{D}_2\rangle$ is orthogonal to the single-excitation superradiant  state $(|\mathrm{eg}\rangle-|\mathrm{ge}\rangle)/\sqrt{2}$. Thus the probe field will not interact with the atom array, resulting in a total transparency point at ${\rm\Delta}_k=0$. Meanwhile, the corresponding inelastic photon flux is zero, i.e., the fluorescence is fully quenched at this point [see Fig.~\ref{inelastic}(a)]. When $N=4$, a total transparency point with zero inelastic photon flux also appears at ${\rm\Delta}_k=0$ [see Fig.~\ref{inelastic}(c)]. the corresponding dark steady state is 

\begin{eqnarray}
|\mathcal{D}_4\rangle&=&\frac{1}{\mathcal{N}_4}\left[3{\rm\Delta}^2|\mathrm{gggg}\rangle+2{\rm\Delta}{\rm\Omega}\left(|\mathrm{eggg}\rangle+|\mathrm{ggge}\rangle-3|\mathrm{gegg}\rangle-3|\mathrm{ggeg}\rangle\right)\right.
\\ \nonumber
&&\left.-4{\rm\Omega}^2\left(|\mathrm{eegg}\rangle+|\mathrm{egeg}\rangle+|\mathrm{gege}\rangle+|\mathrm{ggee}\rangle\right)\right]
\end{eqnarray}
with $\mathcal{N}_4=\sqrt{\left(8{\rm\Omega}^2+{\rm\Delta}^2\right)\left(8{\rm\Omega}^2+9{\rm\Delta}^2\right)}$. Clearly, $|\mathcal{D}_4\rangle$  is 
orthogonal to the superradiant state $(|\mathrm{eggg}\rangle-|\mathrm{gegg}\rangle+|\mathrm{ggeg}\rangle-|\mathrm{ggge}\rangle)/2$. In addition, our numerical calculations show that for larger atom numbers $N=2n~(n\in\mathbb{N}^{+})$, total transmission also appears at  ${\rm\Delta}_k=0$, corresponding to a dark steady state of the system. The transparency phenomenon can be explained as a genuine EIT effect.  

On the contrary, the transmission maxima located at ${\rm\Delta}_k\neq 0$ are not perfect transparency points because the steady state of the driven system is not a dark state.  The fluorescence is not quenched at 
these frequencies, and the corresponding inelastic photon flux is nonzero, as shown by the transmission maxima at ${\rm\Delta}\neq 0$ in Fig.~\ref{inelastic}(b) and \ref{inelastic}(c). Thus around these points, total transmission phenomenon occurs only when a single-photon Fock state incidents, and breaks down outside the single-photon sector.
\section{\label{conclusion}CONCLUSIONS AND DISCUSSIONS}
In summary, we have investigated multiple EIT without a control field in a wQED system containing $N$ atoms. 
By analyzing the collective excitation states of the system and mapping the atom array into a driven $(N+1)$-level system, we provide the physical mechanism of the control-field-free EIT phenomenon in this system.
The EIT-type scattering spectra of the atom-array wQED system are discussed both in the single-photon sector and beyond the single-photon limit. The most significant feather of the multiple EIT scheme discussed here is control-field-free, which may provide an alternative way to produce EIT-type phenomenon in wQED system when external control field is not available. 
The results given in our paper may provide good guidance for future experiments on multiple EIT without a control field in waveguide QED 
system. These results may provide powerful tools for manipulating photon transport in quantum networks, and may have potential applications in multi-wavelength optical communication.
\section{Acknowledgements}
This work was supported by the National Natural Science
Foundation of China (NSFC) under Grants No. 11404269, No.
61871333, and No. 12047576. 

\appendix
\section{\label{Hamiltonian-multi-levelatom}Effective Hamiltonian of a drvien $(N+1)$-level atom coupled to a waveguide}
The configuration of an $(N+1)$-level atom is shown schematically by Fig.~\ref{system}c.  We assume that the transition between the ground state $\left|0\right>$ and the excited state $\left|N\right>$ is coupled by the photon modes in the waveguide, and the excited state  $\left|N\right>$ couples to the metastable state $\left|i\right> (i=1,2,\cdots N-1)$ by a classical laser beam with frequency $\nu_i$ and Rabi frequency ${\rm\Omega}_{i}$, forming a standard driven $(N+1)$-level system that can generate multiple EIT. Under rotating-wave approximation, the Hamiltonian of the system described by Fig.~\ref{system}c can be written as ($\hbar=1$)
\begin{eqnarray}
\hat{H}&=&\tilde\omega_{N}\left|N\right>\left<N\right|+\sum_{i=1}^{N-1}\left[\tilde\omega_{i}\left|i\right>\left<i\right|+\left({\rm\Omega}_{i}e^{-\mathrm{i}\nu_{i}t}\left|N\right>\left<i\right|+\mathrm{H}.\mathrm{c}.\right)\right]
\nonumber \\ &&
+\int\mathrm{d}x\hat{c}_\mathrm{R}^{\dagger}\left(x\right)\left(-iv_{\mathrm{g}}\frac{\partial}{\partial x}\right)\hat{c}_\mathrm{R}\left(x\right)
+\int\mathrm{d}x\hat{c}_\mathrm{L}^{\dagger}\left(x\right)\left(iv_{\mathrm{g}}\frac{\partial}{\partial x}\right)\hat{c}_\mathrm{L}\left(x\right) 
\nonumber \\ &&+\int\mathrm{d}xV_{0N}\delta\left(x\right)\left\{\left[\hat{c}_\mathrm{R}^{\dagger}\left(x\right)+\hat{c}_\mathrm{L}^{\dagger}\left(x\right)\right]\left|0\right>\left<N\right|+\mathrm{H.c.}\right\},
\label{HamiltonianA}
\end{eqnarray}
where $\tilde\omega_i$ represents energy of the level $|i\rangle$. Here we have taken $\tilde\omega_{0}=0$ as a reference. Moving to the interaction picture 
associated with $\hat{H}_{0}=-\sum_{i=1}^{N-1}\nu_i\left|i\right>\left<i\right|$, one can obtain the following Hamiltonian
\begin{eqnarray}
\hat{H}&=&\tilde\omega_{N}\left|N\right>\left<N\right|+\sum_{i=1}^{N-1}\left[\left(\tilde\omega_{N}+{\rm\Delta}_{i}^{(\mathrm{c})}\right)\left|i\right>\left<i\right|+\left({\rm\Omega}_{i}\left|N\right>\left<i\right|+\mathrm{H}.\mathrm{c}.\right)\right]
\nonumber \\
&&+\int\mathrm{d}x\hat{c}_\mathrm{R}^{\dagger}\left(x\right)\left(-iv_{\mathrm{g}}\frac{\partial}{\partial x}\right)\hat{c}_\mathrm{R}\left(x\right)
+\int\mathrm{d}x\hat{c}_\mathrm{L}^{\dagger}\left(x\right)\left(iv_{\mathrm{g}}\frac{\partial}{\partial x}\right)\hat{c}_\mathrm{L}\left(x\right) 
\nonumber \\
&&+\int\mathrm{d}xV_{0N}\delta\left(x\right)\left\{\left[\hat{c}_\mathrm{R}^{\dagger}\left(x\right)+\hat{c}_\mathrm{L}^{\dagger}\left(x\right)\right]\left|0\right>\left<N\right|+\mathrm{H.c.}\right\},
\label{Hamiltonian1}
\end{eqnarray}
where ${\rm\Delta}_{i}^{(\mathrm{c})}=\nu_i-(\tilde\omega_{N}-\tilde\omega_{i})$ is the detuning of the control field coupling the transition $|i\rangle\leftrightarrow|N\rangle$. After tracing out the photon modes in the waveguide, we can obtain the effective non-Hermitian Hamiltonian of the driven $(N+1)$-level atom 
\begin{equation}
\hat{H}=(\tilde\omega_{N}-\frac{\mathrm{i}}{2}{\rm\Gamma}_{N0})\left|N\right>\left<N\right|+\sum_{i=1}^{N-1}\left[\left(\tilde\omega_{N}+{\rm\Delta}_{i}^{(\mathrm{c})}\right)\left|i\right>\left<i\right|+\left({\rm\Omega}_{i}\left|N\right>\left<i\right|+\mathrm{H}.\mathrm{c}.\right)\right],
\label{HeffAtom}
\end{equation}
where ${\rm\Gamma}_{N0}=2V_{N0}^2/v_\mathrm{g}$ is the decay rate from the state $|N\rangle$ to the state $|0\rangle$ into the waveguide modes. Note that in the above derivations, we have neglected the photon loss to the unguided degrees of freedom. 
\section{\label{DegenerateCase} Effective Hamiltonian analysis for the case that some atoms are identical}
In this section, we consider the case that $m_0$ atoms are nonidentical, and among the other $N-m_0$ atoms there are $m_i$ ($i=1,2\cdots M$) atoms with the same frequencies, satisfying $\sum_{i=0}^{M}m_i=N$. One can prove that in this case the number of transparency windows decreases to $m_0+M-1$. To this end we rewrite the Hamiltonian (3) in the matrix form
\begin{eqnarray}
\hat{H}_{\mathrm{eff}}=
\mathbf{\Sigma}^{\dagger}\mathbf{H}\mathbf{\Sigma},
\label{Hdegenerate}
\end{eqnarray}
with $\mathbf{\Sigma}=(\sigma_{1}^{-},\sigma_{2}^{-},\cdots,\sigma_{N}^{-})^{\mathrm{T}}$ and $\mathbf{H}_{ij}=\omega_{i}\delta_{ij}-\mathrm{i}\frac{{\rm\Gamma}}{2}\sum_{i,j=1}^{N}\left (-1\right)^{\left(i-j\right)n}$. Here we let $\omega_1,\omega_2,\cdots\omega_{m_0}$ be the detunings of the $m_0$ nonidentical atoms, and $\omega_{m_0+1}=\cdots=\omega_{m_0+m_1}=\tilde\omega_{m_0+1}$, $\omega_{m_0+m_1+1}=\cdots=\omega_{m_0+m_1+m_2}=\tilde\omega_{m_0+2}$, $\cdots$, $\omega_{\sum_{i=0}^{M-1}m_i+1}=\cdots=\omega_{N}=\tilde\omega_{m_0+M}$.
We further look on the $m_i$ ($i=0,1\cdots M$) atoms as a subsystem, and introduce a unitary transformation 
\begin{equation}
\mathbf{U}={\mathbf{U}}_0\oplus{\mathbf{U}}_1\oplus{\mathbf{U}}_2\oplus\cdots\oplus{\mathbf{U}}_{M}, 
\label{Utrans}
\end{equation}
where 
\begin{equation}
({\mathbf{U}}_i)_{pq}=\frac{1}{\sqrt{m_i}}e^{- \mathrm{i}\frac{2\pi}{m_i}\left(p-1\right)q}\left(-1\right)^{\left(q-1\right)n}.
\label{U}
\end{equation}
After the transformation \eqref{Utrans}, then the Hamiltonian can be written as 
\begin{eqnarray}
\hat{H}_{\mathrm{eff}}=
\mathbf{\Sigma}'^{\dagger}\mathbf{H}'\mathbf{\Sigma}',
\label{Heff5}
\end{eqnarray}
with $\mathbf{\Sigma}'=\mathbf{U}\mathbf{\Sigma}$ and
\begin{eqnarray}
\mathbf{H}'=\mathbf{U}\mathbf{H}\mathbf{U}^{-1}=
\begin{pmatrix}
\mathcal{H}_{1,1}&\mathcal{H}_{1,2}&\cdots&\mathcal{H}_{1,M+1}\\
\mathcal{H}_{2,1}&\mathcal{H}_{2,2}&\cdots&\mathcal{H}_{2,M+1}\\
\vdots&\vdots&\ddots&\vdots\\
\mathcal{H}_{M+1,1}&\mathcal{H}_{M+1,2}&\cdots&\mathcal{H}_{M+1,M+1}.
\\
\end{pmatrix}.
\label{Heffnondegenerate}
\end{eqnarray}
Here
\begin{eqnarray}
\mathcal{H}_{1,1}=\begin{pmatrix}
\tilde\omega_{0}-\mathrm{i}\frac{m_0{\rm\Gamma}}{2}&\lambda_{12}&\cdots&\lambda_{1,m_0}\\
\lambda_{21}&\tilde\omega_{0}&\cdots&\lambda_{2,m_0}\\
\vdots&\vdots&\ddots&\vdots\\
\lambda_{m_0,1}&\lambda_{m_0,2}&\cdots&\tilde\omega_{0}\\
\end{pmatrix},
\label{H11}
\end{eqnarray}
with $\tilde\omega_{0}=\frac{1}{m_0}\sum_{i=1}^{m_0}\omega_{i}$,  $\lambda_{ij}=\frac{1}{m_0}\sum_{s=1}^{m_0}\delta\tilde{\omega}_{s}e^{\mathrm{i}\frac{2\pi}{m_0}(j-i)s}$, and $\delta\tilde{\omega}_{s}=\omega_s-\tilde\omega_{0}$.
\begin{equation}
\mathcal{H}_{i+1,i+1}=\begin{pmatrix}
\tilde\omega_{m_0+i}-\mathrm{i}\frac{m_i{\rm\Gamma}}{2}&0&\cdots&0\\
0&\tilde\omega_{m_0+i}&\cdots&0\\
\vdots&\vdots&\ddots&\vdots\\
0&0&\cdots&\tilde\omega_{m_0+i}\\
\end{pmatrix},
\label{Hii}		
\end{equation}
with $i=1,2,\cdots M$. 
\begin{equation}
\mathcal{H}_{i+1,j+1}=\begin{pmatrix}-\mathrm{i}\sqrt{m_{i}m_{j}}\frac{{\rm\Gamma}}{2}&0&\cdots&0\\0&0&\cdots&0\\
\vdots&\vdots&\ddots&\vdots\\
0&0&\cdots&0\\
\end{pmatrix},
\label{Hij}
\end{equation}
with $i,j=0,1\cdots M$ and $i\neq j$. We can see from above results that each type of $m_i$ ($i=1,2,\cdots M$) identical atoms as a subsystem  can provide a superradiant type collective mode with effective decay $m_i {\rm\Gamma}$ and $m_i-1$ subradiant modes. In addition, these subradiant states not only decouple from the waveguide but also decouple from other states. Therefore, the $N$-atom array is reduced to an effective array containing $m_0+M$ emitters, resulting in $m_0+M-1$ transparency windows.

\bibliography{MS-Oct-1-2022}

\end{document}